\documentclass[12pt]{iopart}

\usepackage{graphicx}
\usepackage{dcolumn}
\usepackage{bm}
\usepackage{float}
\usepackage{lineno}
\usepackage{amssymb}
\usepackage{subcaption}
\usepackage{harvard}
\usepackage{url}

\begin{document}


\title{Training 3D ResNets to Extract BSM Physics Parameters from Simulated Data}

\author{Shawn Dubey$^{1,2}$, Thomas E. Browder$^{1}$, Shahab Kohani$^{1}$, Rusa Mandal$^{3}$, Alexei Sibidanov$^{1}$, Rahul Sinha$^{1}$}

\address{$1$ Department of Physics and Astronomy, University of Hawai‘i at Mānoa,
            2505 Correa Rd., 
            Honolulu, 
            HI 96822,
            USA}

\address{$2$ Center for the Fundamental Physics of the Universe, Department of Physics, Brown University, Department and Organization
            182 Hope St., 
            Providence,
            Rhode Island 02912,
            USA}
            
\address{$3$ Indian Institute of Technology Gandhinagar,
            Department of Physics, 
            Gujarat 382355,
            India}

\begin{abstract}
We report on a novel application of computer vision techniques to extract beyond the Standard Model parameters directly from high energy physics flavor data.  We propose a novel data representation that transforms the angular and kinematic distributions into ``quasi-images", which are used to train a convolutional neural network to perform regression tasks, similar to fitting.  As a proof-of-concept, we train a 34-layer Residual Neural Network to regress on these images and determine information about the Wilson Coefficient $C_{9}$ in Monte Carlo simulations of $B^0 \rightarrow K^{*0}\mu^{+}\mu^{-}$ decays.  The method described here can be generalized and may find applicability across a variety of experiments.
\end{abstract}

\section{Introduction and Motivation}
\label{sec:Introduction}
The decay process 
$B \rightarrow K^{*}\ell^{+}\ell^{-}$ with $\ell = e, \mu$ proceeds via a beauty-to-strange quark ($b\to s$) flavor-changing neutral current, which is forbidden at tree-level in the Standard Model (SM) of particle physics but allowed at second order \cite{PhysRevD.2.1285}.  It is therefore sensitive to beyond the Standard Model (BSM) physics.  

Lepton flavor universality, which is the equal coupling of all lepton flavors ($e$, $\mu$, $\tau$) to the weak interaction is expected to hold in the SM.
A deviation from equal coupling would be a signature of lepton flavor universality violation (LFUV), a BSM physics signature. Several flavor physics experiments have previously measured an apparent deviation from muon-electron universality in a ratio of flavor-specific branching fractions \cite{PhysRevD.86.032012,Aaij2017,PhysRevLett.126.161801}, for $B\to K \ell^- \ell^+$ and $B\to K^* \ell^+ \ell^-$ in the intermediate $q^{2} \equiv M^{2}(\ell^{+}\ell^{-})$ region.  However, LHCb's most recent measurements are consistent with the SM \cite{2212.09152}. 

Nevertheless, there are hints of deviations from the SM in the observed angular distributions of $B\to K^* \ell^+\ell^-$. These can be more clearly identified in angular asymmetries, such as the forward-backward asymmetry ($A_{\mathrm{FB}}$), $S_{5}$, and others, described in Ref. \cite{Aaij2017}. It is possible that these angular asymmetries are lepton-flavor violating and a Belle analysis first directly explored this possibility \cite{PhysRevLett.126.161801}. It is also possible that BSM physics is lepton-flavor universal (LFU).  

Recent LHCb analyses of angular distributions in $B\to K^* \mu^- \mu^+$, including non-factorizable and non-local effects, constrain LFU deviations of the Wilson Coefficient $C_9$ from SM expectations \cite{lhcbcollaboration2024comprehensiveanalysislocalnonlocal,PhysRevD.109.052009}.  Wilson Coefficients (WCs) $C_{i}$ \cite{Schwartz_2013} are parameters of certain effective field theories (EFTs) for the weak interaction of heavy quarks and encode high energy/short distance information.  The WC $C_{9}$ is the coupling of the vector term in these EFTs.
In 2024, results from LHCb fits still indicate a
$\sim 2\sigma$ hint for BSM physics in angular distributions for $C_9$. 
In the future, determining the scenario from which these apparent anomalies originate --- via SM interactions with unaccounted-for hadronic effects, or BSM physics --- is a key experimental problem.  

Rates and binned angular asymmetries in $B\to K^* \ell^+\ell^-$ are measured and used as input to theoretical fitting packages. These packages \cite{Altmannshofer2021,Alguer__2022,HURTH2022136838,ciuchini2022charming}, construct a $\chi^2$ from
the external measurements and are used to determine BSM physics parameters or deviations from the SM Wilson coefficient values.

To circumvent issues associated with conventional fitting methods (discussed in subsequent sections) in extracting BSM parameters, we report on a novel proof-of-concept method that uses a novel data representation and machine learning (ML) to determine BSM physics parameters ($C_{9}$, in our case) directly from data.  A neural network (NN) called a Residual Neural Network (ResNet) \cite{1512.03385}, a variation of the well-known convolutional neural network (CNN) \cite{Geron2019-du} commonly used in computer vision, is employed.  We thus recast the $C_9$ determination problem into a data representation problem optimized for computer vision methods. This recast is a type of data engineering.  

To this end, we employ a Monte Carlo (MC) simulation model \cite{Sibidanov:2022gvb} to generate $B^{0} \rightarrow K^{*0}\mu^{+}\mu^{-}$ events (it is also possible to simulate and study $B \to K^{*0} e^+ e^-$) according to various BSM scenarios, parameterized by the deviations of the WCs from their SM values, $\delta C_{i} \equiv \delta C_{i}^{\mathrm{BSM}} - \delta C_{i}^{\mathrm{SM}}$, where $\delta C_{i} = 0.0$ is the SM case.  From the resulting distributions of the decay variables, we create 3D histograms that are treated as images (``quasi-images" or ``voxel grids"), which are then used to train the ResNet to perform a ML regression task. We restrict ourself to 3D images because 4D (with $q^{2}$ as the fourth dimension) images would have been too computationally intensive on which to generate and regress.

Our approach differs from the more common application of artificial intelligence (AI) methods in HEP physics analyses, which involves classification to distinguish various classes of events such as signal versus background; jet tagging; particle identification (determining particle species); and other classification problems \cite{hepmllivingreview}. In contrast, we will perform regression and
extract a continuous parameter from data. This is similar to ``fitting", a different but essential part of HEP analysis.  We apply NN regression and extract the continuous parameter $\delta C_{9}$.  In other words, we regress on the images that are each generated and labeled with a particular $\delta C_{9}$ value, and use the trained NN regression model to predict $\delta C_{9}$ from images it has never seen previously.

Our NN model learns a mapping between these images and their $\delta C_{9}$ labels, which is equivalent to learning the mapping between the distributions and the $\delta C_{9}$ labels. Our MC result is an example of extracting physics parameters directly from detector data, a desirable result for physics data analyses. We note that the method presented here may be find broad applicability even in the absence of LFV BSM physics.

Other techniques to extract WCs or effective couplings using AI/ML have been used or proposed for LHC experiments. These usually involve template fitting of NN outputs or Bayesian inference \cite{hepmllivingreview,PhysRevLett.121.111801,PhysRevD.98.052004}.
 Our method using the full four-dimensional distribution obtained from the simulated event sample to train a CNN model is different from these approaches. 
 This CNN-based method should make it relatively straightforward to take into account experimental challenges such as backgrounds and detector resolution and does not require projecting down to variables in a lower dimension, which loses potentially discriminating information.
 We also note a related study of a classifier method analogous to a matrix element method, which is used to extract EFT parameters in $W Z$ production \cite{Chen:2020mev}.

\section{Monte Carlo Simulation Model}
\label{sec:MCModel}
We have implemented EFT couplings in a new MC generator \cite{Sibidanov:2022gvb} in the EvtGen framework \cite{Lange:2001uf}.  The new MC generator uses the operator product expansion formalism in terms of the WCs $C_{7}$, $C_{9}$, $C_{10}$, $C_{7}^{\prime}$, $C_{9}^{\prime}$, and $C_{10}^{\prime}$, where the latter three primed WCs correspond to right-handed couplings (the weak couplings in the SM are left-handed) and represent potential BSM physics contributions. 

\begin{figure}
    \centering
    \includegraphics[width=0.8\textwidth]{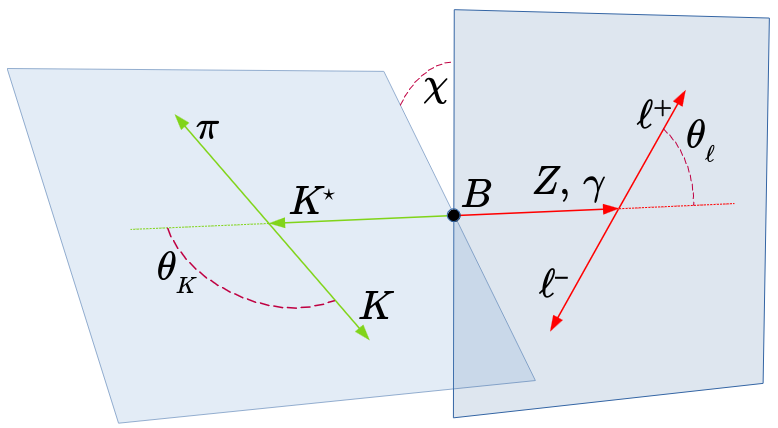}
    \caption{The $B \rightarrow K^{*}\ell^{+}\ell^{-}$ general decay topology showing the observables \protect\cite{Sibidanov:2022gvb}. For this study we only consider the di-muon channel.}
    \label{fig:decay_topology}
\end{figure}

\begin{figure}
\hspace*{-1.6cm}
  \subcaptionbox{$\cos\theta_{\mu}$ Distributions}{%
    \includegraphics[width=0.6\textwidth]{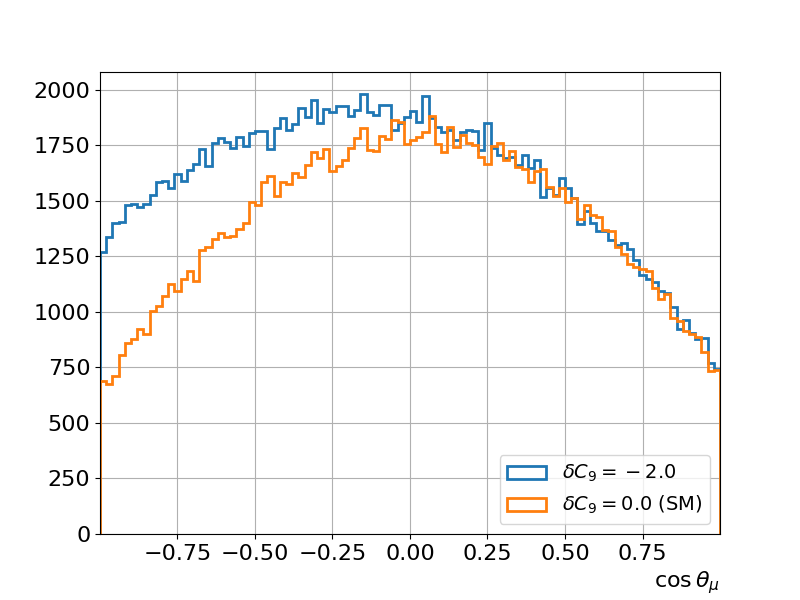}%
  }\qquad
  \subcaptionbox{$\cos\theta_{K}$ Distributions}{%
    \includegraphics[width=0.6\textwidth]{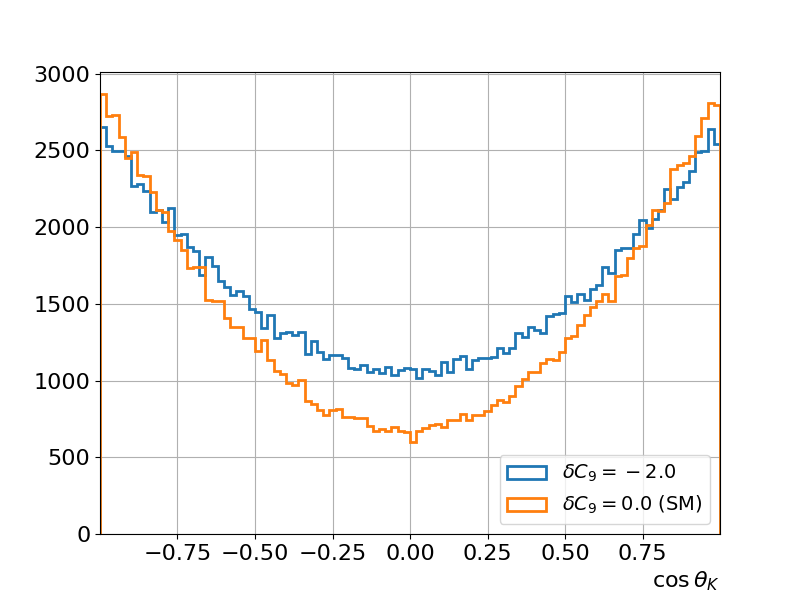}%
  }
  \hspace*{-1.6cm}
    \subcaptionbox{$\chi$ Distributions}{%
    \includegraphics[width=0.6\textwidth]{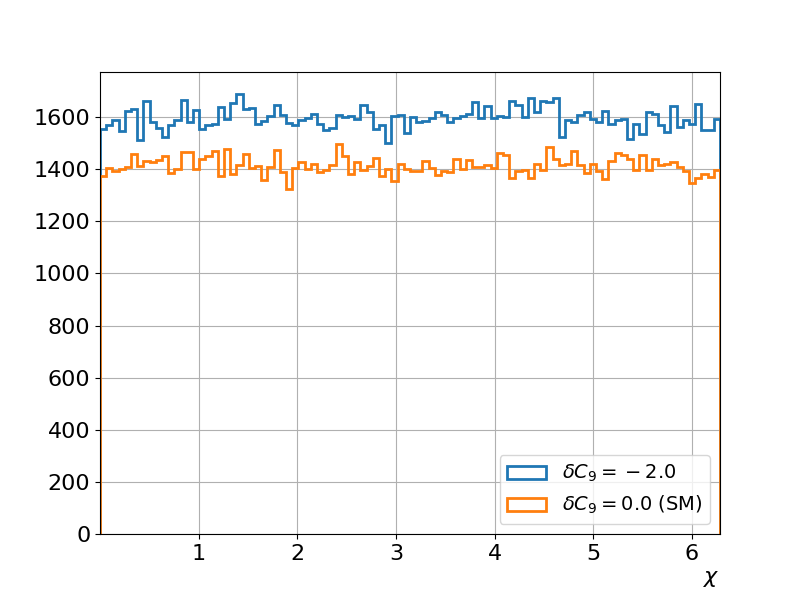}%
  }
    \subcaptionbox{$q^{2}$ Distributions}{%
    \includegraphics[width=0.6\textwidth]{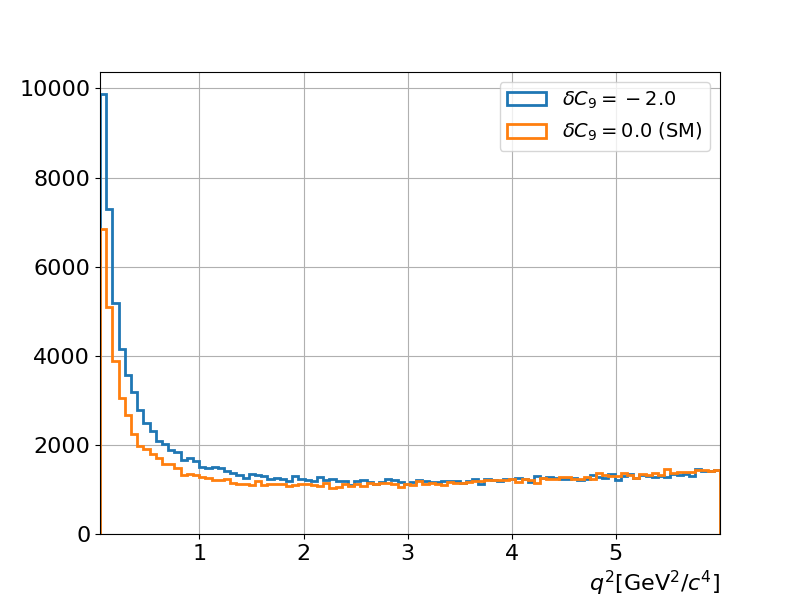}%
  }
    \caption{Distributions of the angular and $q^{2}$ distributions.  Distributions are shown for $1.0 < q^{2} < 6.0$ GeV$^{2}$/c$^{4}$, as that is where the new physics effects are most apparent.  The $q^{2}$ values are binned in bins of the angles to create 3D histograms, i.e. ``quasi-images".  These are input as training instances to the ResNet and treated as images made up of 3D pixels (voxels) in a grid, to train a regression model that learns a mapping between these distributions and $\delta C_{9}$ values.  Two example values are given for each distribution: $\delta C_{9} = 0.0$ (SM) and $\delta C_{9} = -2.0$ (a particular EFT BSM parameterization that is clearly distinguishable from the SM case).}
  \label{fig:vardists}
\end{figure}

Our EvtGen model is parameterized in terms of the WCs' deviation from their SM values, $\delta C_{i}$.  Each of the $\delta C_{i}$ can be chosen by the user \cite{Sibidanov:2022gvb}. Choosing a non-zero $\delta C_i$ has the effect of altering the correlations between four variables: $q^{2}$, defined previously, the cosine of the lepton helicity angle $\cos\theta_{\ell}$ ($\ell = \mu$ in our case), the cosine of the helicity angle of the $K$, $\cos\theta_{K}$, and the angle $\chi$ between the decay planes of the di-lepton and $K^{*}$ decay planes.  Figure \ref{fig:decay_topology} shows the general decay topology and the full set of angular observables while Figure \ref{fig:vardists} shows examples of their distributions for two different values of $\delta C_{9}$.  Distributions are shown for $1.0 < q^{2} < 6.0$ GeV$^{2}$/c$^{4}$, where the new physics effects are most apparent.

\section{Creating the Images}
\label{sec:images}
As described in Sec. \ref{sec:MCModel}, we employ an MC event generator to produce $B^{0} \rightarrow K^{*0}\mu^{+}\mu^{-}$ events.  For simplicity, we only consider the $C_{9}$ WC, the coupling of the vector term of the model implemented in Ref. \cite{Sibidanov:2022gvb}.  While we focus on the region where $\delta C_{9} \leq 0.0$, as theory fits seem to favor a negative $\delta C_{9}$ near $-0.9$ for the di-muon channel \cite{Altmannshofer2021}, samples of MC events are generated with $\delta C_{9} \in [-2.0, 1.1]$.  This allows the NN to better learn the physics distributions near $\delta C_{9} = 0.0$.  Twenty-two $\delta C_{9}$ values are chosen in the above range and $5\times10^{5}$ $B^{0} \rightarrow K^{*0}\mu^{+}\mu^{-}$ events are generated for each.  Only MC events for $B^{0} \rightarrow K^{*0}\mu^{+}\mu^{-}$ are used and not charge conjugate events, as some of the angular asymmetries inherent in the angular distributions that allow for discrimination between SM and BSM physics would cancel if the charge conjugated events are included in the same images.

As this is a proof-of-concept study, (quasi-)images are produced with high statistics to clearly demonstrate our ML method is capable of learning subtle features in them.  Approximately $2.4\times10^{4}$ $B^{0} \rightarrow K^{*0}\mu^{+}\mu^{-}$ events populate a single image. This number of events corresponds to approximately 250 ab$^{-1}$-equivalent integrated luminosity at a Belle II upgrade (assuming the Belle signal reconstruction efficiency of approximately 25\%, given in \cite{PhysRevLett.126.161801}), five times the Belle II target integrated luminosity \footnote{This method can be extended to smaller datasets, for example, the 50 ab$^{-1}$ data sample from Belle II.}.  Therefore the quotient of $5\times 10^{5}$/$2.4\times 10^{4}$ is the number of images produced for each $\delta C_{9}$ value in a single image generation campaign.  We perform multiple campaigns in order to have sufficient statistics for training and testing the ResNet.  

\begin{figure}
\hspace*{-2.5cm}
  \subcaptionbox{Voxel grid image for $\delta C_{9} = -2.0$}{%
    \includegraphics[width=0.64\textwidth]{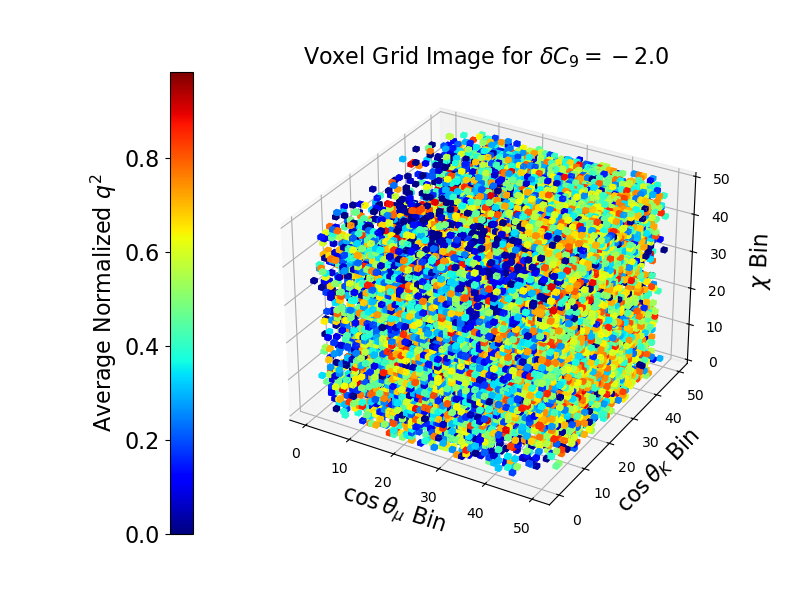}%
  }\qquad
  \subcaptionbox{Voxel grid image for $\delta C_{9} = 0.0$ (SM)}{%
    \includegraphics[width=0.64\textwidth]{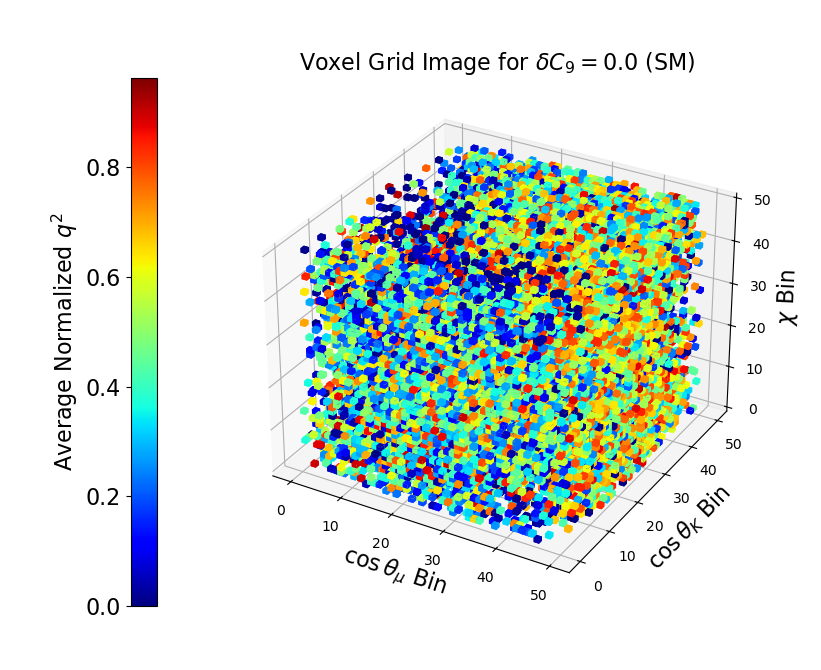}%
  }
    \caption{Voxel grid images (``quasi-images") used for training and evaluation of the ResNet.  The average, normalized, $q^{2}$ values are binned into 50 equal-width bins in the angular variables.  This provides enough resolution and detail in the images while maintaining a reasonable amount of computational resources needed for training.  Examples for the cases of $\delta C_{9} = 0.0$ (SM) and $\delta C_{9} = -2.0$ are shown.  The color of the voxels indicates only the average, normalized, $q^{2}$ value in a 3D bin.} 
  \label{fig:voxelgrids}
\end{figure}

To create the images, we find the average, normalized\footnote{The normalization allows us to put the $q^{2}$ values on the same scale as the values of the angular values, which can aid in model training \cite{Geron2019-du}}, $q^{2}$ values over all events and bin these values in 50 equal-width bins in the three angular variables.  The entire range of $q^{2}$ for $q^{2} \leq 20.0$ GeV$^{2}$/c$^{4}$ is used.  
Bins with no events have a value of 0.0.  Therefore, the shape of the input image is (height, width, depth) = (50, 50, 50).  In effect, we have created a grid of voxels (3D pixels).  The images are treated as ``grayscale", so that the images are input as tensors to the neural network and each have shape (50, 50, 50, 1), where the value 1 denotes the number of channels.  Figure \ref{fig:voxelgrids} shows two examples of the signal images for different values of $\delta C_9$.

We note that we produce only generator-level MC signal events to populate the images, i.e. we do not yet include backgrounds, efficiencies, or detector resolution.  In the future, for Belle II-specific studies, data-driven backgrounds will be obtained from the beam-energy-constrained mass sidebands of the signal region.  The beam-energy-constrained mass is defined as $M_\mathrm{bc} = \sqrt{E^{2}_{\mathrm{beam}}/c^{4} - p^{2}_{B}/c^{2}}$, where $E_{\mathrm{beam}} = \sqrt{s}/2$ and $p_{B}$ is the magnitude of the $B$-meson three-momentum in the center-of-mass frame of the colliding $e^+e^-$ beams.

\section{The Neural Network}
As discussed above, a ResNet variation of the CNN is employed.  ResNets were developed
to better train deep neural networks and can help solve problems such as vanishing gradients in deep neural networks \footnote{A deep neural network here is one with more than two hidden layers of trainable parameters.}, in which the gradient shrinks during backpropagation, making it difficult for the network to learn \cite{Geron2019-du}.  ResNets solve this problem by introduction of a ``skip" or ``shortcut" connection.  These connections send the signal that comes into a layer directly to a layer further up the network.  This forces the model to learn the residual between the input and the underlying function mapping, rather than the function mapping itself.  The ResNet architecture was chosen as it is known to perform well on many computer vision tasks; our previous studies using fully-connected neural networks and simpler CNNs did not perform well.

\begin{figure}
    \centering
    \includegraphics[width=0.8\textwidth]{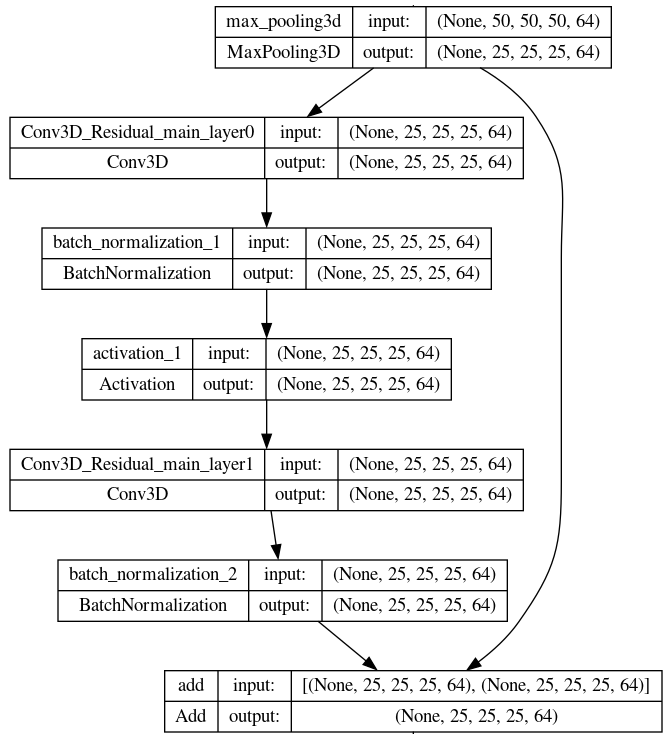}
    \caption{A portion of the topology showing some of the layers and a skip connection in the ResNet, produced by $\texttt{Keras}$.  The skip connection feeds the input directly into the output and adds them. This prompts the NN to learn the function mapping residual and may help mitigate issues with deep neural network training, such as vanishing gradients.}
    \label{fig:flowchart}
\end{figure}

Specifically, a three-dimensional variation of the ResNet introduced in Ref. \cite{1512.03385} is used and built using $\texttt{TensorFlow 2.17.0}$ \cite{tensorflow2015-whitepaper} and $\texttt{Keras 3.4.1}$ \cite{chollet2015keras}.  At the bottom of the network, there are an input layer; a convolutional layer with 64 filters, a kernel size of 7, and a stride of 2; a batch normalization layer followed by a ReLU activation function; and a max pooling layer with a pool size of 3 and a stride of 2.  There are 34 convolutional layers in the main NN path, utilizing the ReLU activation function \cite{Geron2019-du} and $\texttt{same}$ padding.  Stochastic gradient descent is used for optimization.  The loss function is the mean absolute error (MAE), given by $\frac{1}{n}\sum_{i=1}^{n}|y_{i}^{\prime} - y_{i}|$, where $y_{i}^{\prime}$ is the predicted value, $y_{i}$ is the true value, and $n$ is the number of instances \cite{Geron2019-du}.  At the top of network there is a global average pooling layer; a flatten layer; and one fully-connected (dense) layer employing 1000 neurons, followed by a dropout layer with a 50\% dropout probability.  The final layer is a dense layer with one neuron and a linear activation function that performs the regression task to extract $\delta C_{9}$ values directly from image test set \footnote{The linear activation function is used as its range is a continuum of values $\in$ ($-\infty$, $\infty$).}. No hyperparameter optimization was done as the network performance initially worked well.

A topology chart, provided by $\texttt{Keras}$, of part of the ResNet implemented here is shown in Fig. \ref{fig:flowchart}.  Each cell in the flow chart has two rows: the top row contains the layer label on the left and the input shape of the input tensor on the right.  The bottom row contains the type of layer and the output shape of the tensor.  The shapes are given in Python form, where the ``None" value means the layer can operate on a batch size that can change.  The next three values are the height, width, and depth.  Descriptions of the different type of layers used and their effect on the tensors and their dimensions are given in GitLab repository at \cite{Geron2019-du}.

\section{Training the Neural Network} 
Voxel grid images are generated for $\delta C_{9} \in$ [-2.0, 1.1].  Due to different MC generation campaigns, the number of voxel grid images per $\delta C_{9}$ value is unbalanced.  For $\delta C_{9} \in$ [-2.0, 0.0], the number of voxel grid images is 2640 in the training set and 720 for the validation set, except for $\delta C_{9} = -0.9$ and $\delta C_{9} = -0.4$, for which there are 5280 voxel grid images in the training set and 1440 in the validation set.  For $\delta C_{9} \in$ [0.1, 1.1] there are 2340 voxel grid images in the training set and 540 in the validation set.  A batch size of 128 voxel grid images is used during training.  The test set contains 900 images per $\delta C_{9}$ value.

To facilitate learning, the learning rate is reduced every five epochs by a factor of $1/5$, if no improvement in validation loss is seen, with a minimum loss of $10^{-6}$ possible. Early stopping is implemented if there is no improvement in the validation loss (MAE) after 20 epochs, with the maximum number of training epochs set to 1000.  Training is then performed using the GPU nodes of the University of Hawai‘i's Koa HPC cluster, using the $\texttt{Slurm}$ job scheduler \cite{SchedMD}.  As each images has a $\delta C_{9}$ label associated with it, this is a supervised learning task.  The relevant code for the NN, MC generation, voxel grid generation and visualization, training, test set evaluation may currently be found in Ref. \cite{btokstarll_code}.

\section{Results}
\begin{figure}
\hspace*{-2.5cm}
  \subcaptionbox{Voxel grid image for $\delta C_{9} = 0.0$ (SM)}{%
    \includegraphics[width=0.65\textwidth]{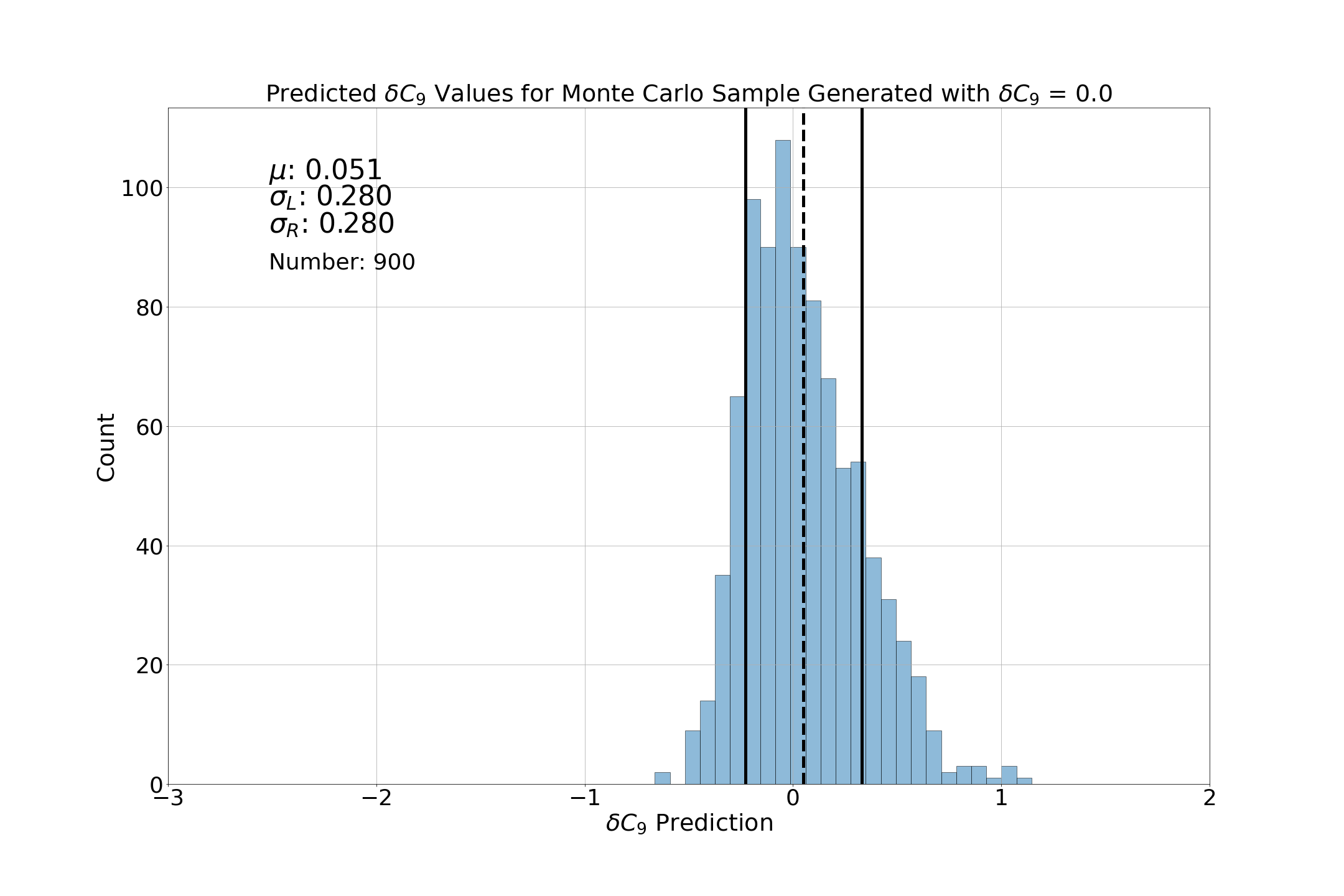}
  }\qquad
  \subcaptionbox{Voxel grid image for $\delta C_{9} = -2.0$}{%
    \includegraphics[width=0.65\textwidth]{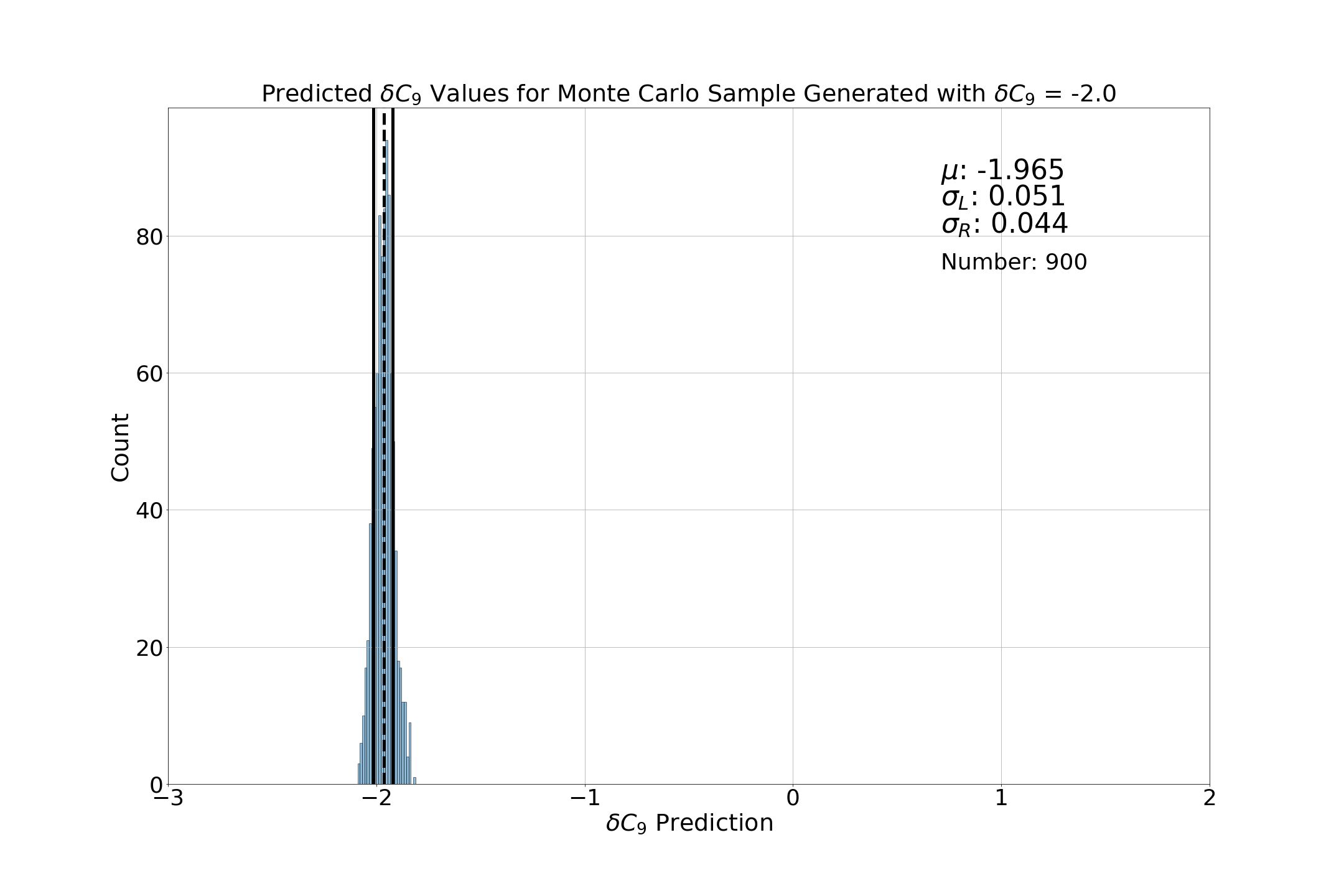}
  }
    \caption{Predictions for an ensemble of 900 independent images that were reserved for testing, for two specific $\delta C_{9}$ values ($\delta C_{9}$ = $0.0$ and $-2.0$). The mean (black dashed line) and left ($\sigma_{L}$) and right ($\sigma_{R}$) standard deviations (black solid lines) are then determined from the distribution.  The horizontal axis is fixed to the same range in both figures to demonstrate that the distributions closer to the SM value of $\delta C_{9}$ are wider due to the images being harder for the ResNet to distinguish, given the small BSM contributions in this region.}
  \label{fig:ensembles}
\end{figure}

As this is a AI/ML model for regression and not classification, standard tools to assess the trained model used in classification, e.g. the Receiver Operating Characteristic (ROC) curve \cite{Geron2019-du}, are not applicable here.  Instead for the test set, we utilize ensembles of MC simulation experiments.  For each of the 22 WCs that were used to generate training images, we generate an ensemble of statistically independent samples of 900 images.  Each of the ensembles of 900 images are evaluated by the trained network and distributions of predicted $\delta C_{9}$ values are obtained.  We also generate 900 images using $\delta C_{9}$ values that are between the ones used to generate the training set images.  This provides a further test of model robustness that tests the model's ability to predict $\delta C_{9}$ values for images for which the model has not been explicitly trained.  Examples of these prediction distributions are provided in Fig. \ref{fig:ensembles}.  Using the Pandas software \cite{reback2020pandas,mckinney-proc-scipy-2010}, these distributions are used to calculate a mean ($\mu$), and a left ($\sigma_{L}$) and right ($\sigma_{R}$) standard deviation, determined from the points where the cumulative distribution function is 16\% and 84\%, respectively. 

\begin{figure}
  \centering
  \includegraphics[width=0.8\textwidth]{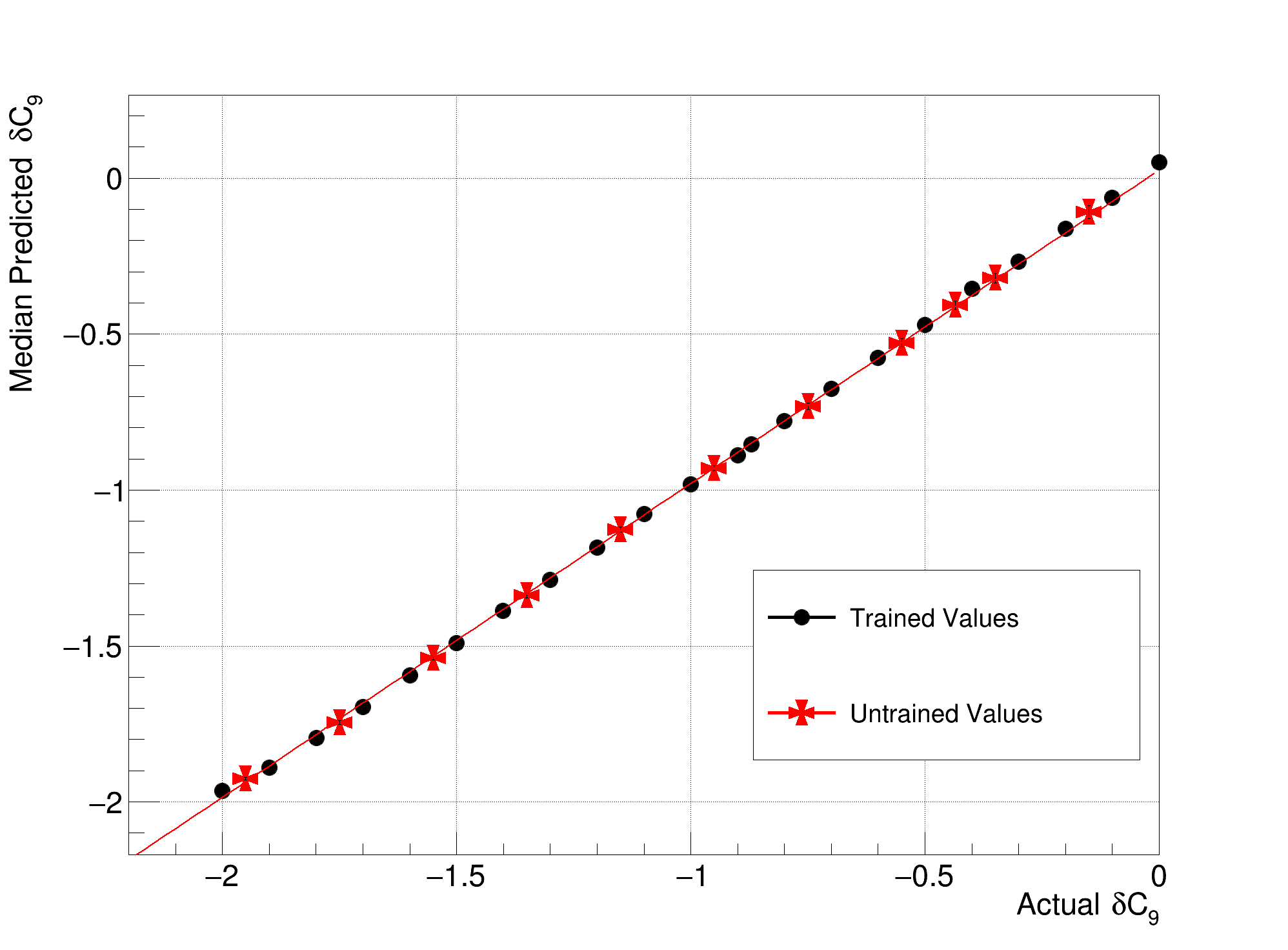}
  \caption{Linearity verification from the MC ensemble experiments.  The mean predicted $\delta C_{9}$ values are plotted against the actual (generated) $\delta C_{9}$ value.  Black circles are from MC ensemble experiments in which images are generated using $\delta C_{9}$ values that the model had seen in training.  Red crosses are from MC ensemble experiments in which images are generated using $\delta C_{9}$ values that the model had not seen in training.  The error bars are scaled by $\sqrt{\mathrm{N}}$, where $\mathrm{N}$ is the number of test images in the MC ensemble for each $\delta C_{9}$ value ($\mathrm{N}$ = 900), and are thus too small to be visible in this figure.}
  \label{fig:linearity2}
\end{figure}

\begin{figure}
  \centering
  \includegraphics[width=0.8\textwidth]{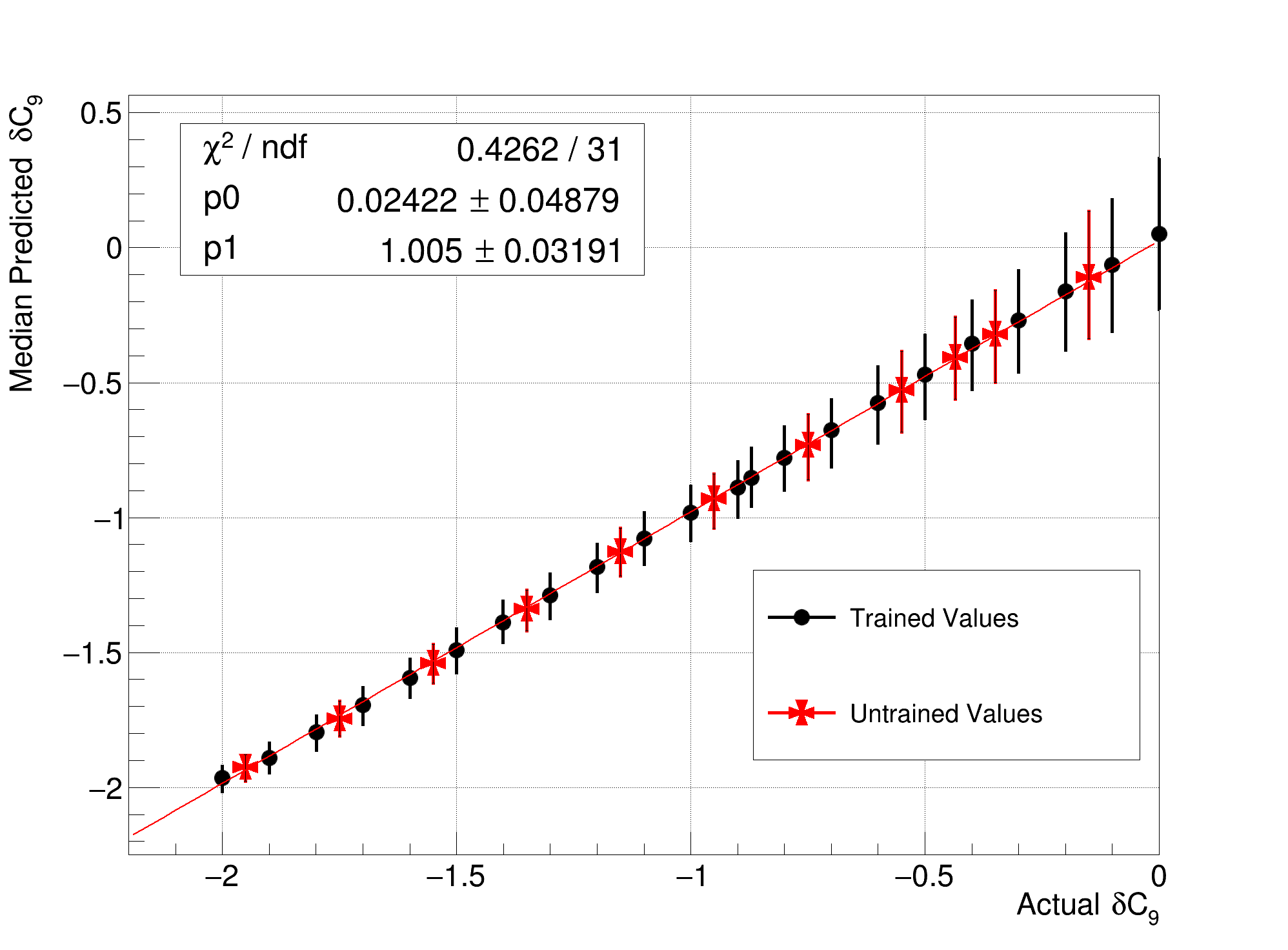}
  \caption{Linearity test from the MC ensemble experiments illustrating the sensitivity at each $\delta C_{9}$ value.  The mean predicted $\delta C_{9}$ values are plotted against the actual (generated) $\delta C_{9}$ value.  Black circles are from MC ensemble experiments in which images are generated using $\delta C_{9}$ values that the model had seen in training.  Red crosses are from MC ensemble experiments in which images are generated using $\delta C_{9}$ values that the model had not seen in training.  In contrast to Fig. \ref{fig:linearity2}, the error bars are not scaled by $\sqrt{\mathrm{N}}$, where $\mathrm{N}$ is the number of test images in the MC ensemble for each $\delta C_{9}$ value ($\mathrm{N}$ = 900). Here the left ($\sigma_{L}$) and right ($\sigma_{R}$) standard deviations are determined from the points where the cumulative distribution functions of the results of the ensemble experiments are 16\% and 84\%, respectively.}
  \label{fig:linearity}
\end{figure}

  The predicted results from the NN are then plotted against their actual (generated) values to obtain a linearity plot, given in Fig. \ref{fig:linearity2}.  In the linearity test, to take into account the correlations between the generated samples, the error bars are scaled down by $\sqrt{\mathrm{N}}$, where $\mathrm{N}$ is the number of test images in the MC ensemble for each $\delta C_{9}$ value ($\mathrm{N}$ = 900).  Figure \ref{fig:linearity} shows the same points with their unscaled error bars, which gives the statistical sensitivity based on the NN and the test set MC ensemble.  As global theory fits and LHC results for the dimuon-specific $\delta C_{9}$ appear to favor a negative value near -0.9 \cite{Altmannshofer2021}, we consider only values for $\delta C_{9} \leq 0.0$ in Fig. \ref{fig:linearity2} and Fig. \ref{fig:linearity}. 

\section{Discussion}
As seen in the linearity plot in Fig. \ref{fig:linearity}, the ResNet appears to be able to obtain a mapping between $\delta C_{9}$ values and MC signal events when those events are recast into images.  However, as one approaches the SM $\delta C_{9} = 0.0$ value, the uncertainties increase. The images closer to the SM are more similar and difficult to distinguish by the ResNet; we note that for $\delta C_9$ near zero, the BSM contributions are smaller.  The uncertainty in the ResNet predictions is therefore larger toward $\delta C_{9} = 0.0$.  When compared to the results of the 4D unbinned maximum likelihood fit to generator-level signal MC samples in Ref. \cite{Sibidanov:2022gvb}, the upper error bar ($\sigma_{R}$) from the ensemble test for $\delta C_{9} = 0.0$ is larger, at 250 ab$^{-1}$-equivalent signal events.  Nevertheless, we do not interpret this as a major defect for the method presented here.  The sensitivity is 
likely to be ameliorated with more training data.  

Further, using a 4D unbinned maximum likelihood fit has a number of problems in a real high energy physics experiment.  In the presence of backgrounds and detector resolution, it is difficult to parameterize the backgrounds, efficiency, and resolution in multiple dimensions.  The associated issues may be greatly mitigated if the problem is recast as a computer vision problem, as none of these parameterizations are required  in the NN approach. 

The main issue with the method described here is one of computational power and storage.  We used $\delta C_{9}$ and  high-statistics generator level MC samples for the proof-of-concept demonstration.  In reality, a fully trained and useful model would have to be trained using images created according to all the WCs mentioned above, as well as different integrated luminosities (assuming applicability at Belle II) and appropriate backgrounds. 

As an initial naive approach, prior to this study, we attempted to train a fully-connected neural network \cite{Geron2019-du} using one $B$ ($\bar{B}$) decay event as one training instance, with the associated angular and $q^{2}$ values as training features. This approach did not yield a NN that was able to learn to map the kinematic input values to a $\delta C_{9}$ output value.  This is likely due to the fact that such a NN has difficulty finding correlations on an event-by-event basis due to subtle differences in the distributions.  Thus recasting the problem as one of building a data representation that is effective for computer vision appears to be effective: many events are used to create a single image (i.e. a single training instance), and having a large number of events in a single training instance makes the correlations between the angular and $q^{2}$ distributions, and $\delta C_{9}$ values, more apparent.

\section{Conclusion}
We have trained a three-dimensional ResNet to learn a mapping between different $\delta C_{9}$ values and images, where the images are created using kinematic distributions obtained from MC simulations of $B^0\to K^{*0} \mu^+ \mu^-$ decays\footnote{It is also possible to perform this study with simulations of $B^0\to K^{*0} e^+ e^-$ events}.  We have recast the problem of fitting complicated multi-dimensional distributions using maximum likelihood techniques to one of developing a data representation for computer vision.  This should make it relatively easy to take into account experimental complexities such as backgrounds and experimental resolutions and does not require projecting down to a lower dimension (e.g. in this case to angular asymmetries such as $A_{\mathrm{FB}}$ and $S_{5}$), losing potentially valuable information.  Work is on-going to include efficiencies, detector resolutions, and backgrounds that will be obtained from beam-energy-constrained mass sidebands in real experimental data. 
Our approach may also find application to studies of $\bar{B}^{0} \rightarrow D^{\ast +} \ell^{-} \bar{\nu}$, where a new BSM physics generator has also recently been developed \cite{PhysRevD.107.015011}.  
These results can also be extended to the determination of multiple BSM Wilson coefficients e.g. $C_7$, $C_10$, and right-handed primed couplings.

We have shown that a ResNet is able to learn a mapping and successfully extract information about an important physics parameter.  Difficulties with this method will likely be mitigated with increased training set sizes and additional computational resources.  

This CNN-based regression method is quite general.  While we have performed our study with an eye toward Belle II and flavor physics experiments, it should be applicable to other areas of high energy physics. For example, it may be useful in the study of $t\bar{t}H$ or $W W Z$ couplings at CMS and ATLAS,  which has already commenced using machine learning approaches \cite{Tonon2021}.  In general, we hope to provide a more efficient and flexible method of performing BSM physics searches that are broadly applicable across different fields of physics.

A preliminary version of the results described above were presented in a poster during the 2023 Computing in High Energy Physics conference in Norfolk, Virginia and ICHEP 2024 in Prague, Czech Republic, and are briefly discussed in their Proceedings. This work supersedes earlier results.

\section{Acknowledgments}
We thank our colleagues on Belle II as well as the KEK computing group for their excellent operation of the KEK computing center.  We have used the University of Hawai‘i MANA HPC cluster.  The technical support and advanced computing resources from University of Hawai‘i Information Technology Services – Cyberinfrastructure, funded in part by the National Science Foundation CC* awards \# 2201428 and \# 2232862 are gratefully acknowledged.  We also thank Hongyang Gao (ISU) for his seminal suggestion to use computer vision techniques to search for new physics couplings, and Chunhui Chen (ISU) for facilitating the meeting, as well as Peter Sadowski (UHM), Jeffrey Schueler (UNM) and Sven Vahsen (UHM) for their helpful discussions on machine learning and technical advice.  We acknowledge Ethan Lee (UHM) for cross-checks and investigation of applicability of this method with full GEANT4 Monte Carlo simulations.

T.E.B., S.D., S.K., and A.S. thank the DOE Office of High Energy
Physics for support through DOE grant DE-SC0010504.

\section{References}

\bibliographystyle{jphysicsB} 
\bibliography{references}

\end{document}